\pdfoutput=1

\documentclass[11pt, oneside]{article}   	% use "amsart" instead of "article" for AMSLaTeX format
\usepackage{geometry}                		% See geometry.pdf to learn the layout options. There are lots.
\usepackage{graphicx}				% Use pdf, png, jpg, or eps§ with pdflatex; use eps in DVI mode
\usepackage{amsmath}				% TeX will automatically convert eps --> pdf in pdflatex		
\usepackage{amssymb, bm}

\usepackage[square,sort,comma,numbers]{natbib}
\usepackage{hyperref}
\usepackage{mathtools}
\usepackage{multirow}
\usepackage{lscape}
\usepackage[skip=10pt]{caption}

\newcommand{\comment}[1]{}

\begin{document}

\title{Bayesian Sequentially Monitored Multi-arm Experiments with Multiple Comparison Adjustments}
\author{Andrew W. Correia\\
SessionM, Inc.\\
2 Seaport Lane, 11$^{th}$ Floor\\ Boston, MA 02210\\ 
%\email{acorreia@sessionm.com}}
\href{mailto:acorreia@sessionm.com}{acorreia@sessionm.com}
\date{}}

\maketitle

\begin{abstract}
Randomized experiments play a major role in data-driven decision making across many different fields and disciplines. In medicine, for example, randomized controlled trials (RCTs) are the backbone of clinical trial methodology for testing the efficacy of new drugs and therapies versus existing treatments or placebo. In business and marketing, randomized experiments are typically referred to as A/B tests when there are only two arms, or variants, in the experiment, and as multivariate A/B tests when there are more than two arms. Typical applications of A/B tests include comparing the effectiveness of different ad campaigns, evaluating how people respond to different website layouts, or comparing different customer subpopulations to each other.

This paper focuses on multivariate A/B testing from a digital marketing perspective, and presents a method for the sequential monitoring of such experiments while accounting for the issue of multiple comparisons. In adapting and combining the methods of two previous works, the method presented herein is straightforward to implement using standard statistical software and performs quite well in various simulation studies, exhibiting better power and smaller average sample sizes than comparable methods.
\end{abstract}

\section*{Introduction}
In traditional A/B testing, a researcher conducts a randomized experiment with two arms, or variants, A and B, which can be thought of as the control and treatment group, respectively. The objective of the experiment is to determine if the people randomized to group B performed better towards some outcome or goal than did the people randomized to the control group, A. One simple example of this would be a clinical trial for patients with high blood pressure testing the efficacy of some new high blood pressure medication versus placebo. Researchers would be interested in knowing whether the group receiving the new medication, group B, saw its members experience a greater reduction in blood pressure levels than did the placebo group, group A. In perhaps a more familiar marketing setting, a typical example might be assigning one group of people, group B, to be shown a certain advertisement while the control group, group A, is not shown that ad. The marketer might then be interested in, say, knowing if group B exhibited an increase in some type of purchasing behavior compared to the control group, group A.
\\

A multivariate A/B test extends the standard A/B test to include more than one treatment group, but still has only one control group. Let the control group continue to be denoted as group A, then the different arms, or variants, in a multivariate A/B test can be thought of as: A, $\mathrm{B}_{i}$, $i = 1, 2, ..., m$, where $m$ is some arbitrary number of variants (excluding the control group) to be included in the experiment. As before, an example of a multivariate A/B test might be the same clinical trial for high blood pressure medication, but with three new candidate treatments each being compared against baseline, instead of just one.\footnote{Throughout this paper, baseline and control will be used interchangeably and are taken to mean the same thing. The baseline success rate, for example, is equal to the success rate in the control group.} Similarly, in the context of marketing, a multivariate A/B test would look very much like the previous marketing example, but with, say, four or five different creative types each being compared against the control group, A.
\\

When there are multiple arms in an experiment there are a number of questions a researcher might hope to answer. Some of those questions could include the following:

\begin{itemize}

\item What is the best performing variant?

\item What are the top three performing variants?

\item From a statistical testing standpoint, is each variant significantly better than baseline?

\item How do the variants compare to each other? 

\item For each variant, what is the probability that it is the best performing variant?

\end{itemize}

The above questions come in addition to the usual questions that arise when undertaking a classical two-arm A/B test, such as:

\begin{itemize}

\item How many people do I need in each group to detect a statistically significant difference?

\item When I'm calculating my sample size, how do I know how big of a difference I want to detect?

\item Can I stop as soon as there's a significant difference between groups?

\item How soon can I look at the data from this experiment?

\end{itemize}

In the rest of this paper, I outline my proposed methodology for multivariate A/B testing and how it addresses each of the questions outlined above. I begin with a brief overview of statistical hypothesis testing and the different study designs that can be implemented in hypothesis testing under a simple two-arm A/B test. I then present the proposed multivariate A/B testing methodology followed by results for that methodology applied to simulated multivariate A/B experiments, and close with a discussion of possible extensions and additions to this methodology.

\section*{Methodology}
\subsection*{Overview}

At its core, a standard two-arm A/B test is a simple statistical hypothesis test. Denote the null hypothesis by $H_{0}$ and the alternative hypothesis by $H_{1}$. The hypotheses being tested then are typically either:

\begin{enumerate}

\item $H_{0}:$ Groups A and B perform identically towards the same goal of interest\\
vs.\\
$H_{1}:$ Groups A and B do not perform identically towards the same goal of interest,

or

\item $H_{0}:$ Group A performs no worse than group B towards the same goal of interest\\ 
vs.\\ 
$H_{1}:$ Group B performs better than group A towards the same goal of interest.

\end{enumerate}

For clarity, assume a digital marketer is interested in converting customers to the latest upgrade of his company's product. To do this, he wants to launch a new campaign aimed at generating conversions. But before launching this campaign to all customers he wants to be sure it will work, so he first decides to conduct an A/B test wherein some customers are randomly assigned to see the new ad (group B) and some are not (group A). Let $\mu_{A}$ denote the conversion rate in the control group, A, and $\mu_{B}$ the conversion rate in the exposed group, B. Then the marketer's A/B test will look to test either:

\begin{enumerate}

\item $H_{0}: \mu_{A} = \mu_{B}$ vs. $H_{1}: \mu_{A} \neq \mu_{B}$

or

\item $H_{0}: \mu_{A} \geq \mu_{B}$ vs. $H_{1}: \mu_{A} < \mu_{B}$

\end{enumerate}

To formally test either of these hypotheses there are a few analytical approaches to consider. First note that in the marketing example provided, customer conversion rates are being analyzed. This means that the outcome variable is dichotomous (yes or no, 0 or 1, etc.). For each individual in the experiment, the outcome variable of interest is the variable capturing whether this person installed the latest upgrade, yes or no. To compare the two groups, the analysis could take the form of a Z-test for the comparison of two proportions (assuming the sample size in each group is large enough), Fisher's exact test for the comparison of two proportions, a simple Bayesian analysis comparing the posterior distributions of two Beta distributions, or if there are other variables that should be considered in the model then perhaps a logistic regression wherein a statistical test is conducted on the coefficient of interest via, e.g., a Wald test or a likelihood ratio test, etc.

\subsection*{Study Design}

In addition to the analytic method, the researcher also needs to decide on a study design. The study design could take the form of a fixed-sample size design, or a sequential sample size design (with or without adaptive randomization) \cite[Ch. 87]{trials_handbook}. In a fixed-sample design, the number of individuals, or customers, to be enrolled in the experiment is fixed \emph{a priori} based on a sample size calculation prior to the start of the experiment \cite[Ch. 87]{trials_handbook}. That calculation will depend on a number of factors, including: an estimate of the treatment effect (i.e., how much better/worse the exposed group will perform compared to the control group), an estimate of the baseline success (conversion) rate in the control group, the desired Type I error rate for the test ($\alpha$), the desired power to detect the estimated treatment effect (power = 1 - Type II error rate), and the proportion of individuals to be randomized to each group (e.g. 50\% in each group, 60\%/40\%, etc) \cite{fda_adap}. In a fixed-sample size design, no conclusions regarding the hypotheses being tested can be drawn until the full sample has been accrued in each group, otherwise the Type I error will be greatly inflated beyond the pre-specified level \cite{armitage69, mehta}. Additionally, the proportion of individuals randomized to each arm of the experiment will typically remain fixed throughout the experiment. The fixed design is often referred to as a traditional \cite{mehta} or conventional \cite{fda_adap} study design, and it is how most experimental trials are carried out \cite[Ch. 87]{trials_handbook}.
\\

A sequential sample size design, sometimes also called an adaptive \cite{fda_adap} or flexible design \cite{mehta}, will monitor the data accrued throughout the experiment either continuously, or at administratively convenient intervals in order to make decisions about the future course of the study while the study is underway \cite{mehta}. At each interim point, a decision can be made regarding whether the experiment should be stopped because the treatment arm has shown itself to be either a clear winner or loser with respect to the control group. If there are multiple treatment arms in the study, decisions can also be made regarding the re-allocation of individuals to different arms of the experiment by altering the assignment probabilities so that more people are randomized to better performing arms as the study is ongoing, or by completely dropping arms that are exhibiting a very high probability that they are ineffective.\footnote{Note that throughout the rest of this document, unless otherwise stated, the term adaptive will only be used with respect to adaptive randomization, i.e. continuously updating the randomization of individuals to treatment groups based on how well each group has performed up until that point - better performing groups would receive more subjects, poorly performing groups fewer. This is known as outcome-adaptive randomization \cite[Ch. 88]{trials_handbook}. More on this in the Discussion section.} Under such a design, the sample size of the study is typically not determined and fixed \emph{a priori}, though in many cases a maximum sample size will be defined \emph{a priori} in order to keep the size of the experiment manageable. Similarly, in some cases the postulated treatment effect need not be provided as well. However the desired power and Type I error rate will play a role in study design, and adjustments need to be made to classical test statistics to avoid inflating the Type I error of the test. For studies in which the experimenter has pre-determined when and how many interim looks at the data there will be, a common frequentist approach to controlling the Type I error of a sequentially-designed study is to specify an alpha spending function \cite{lan_demets}, examples of which include the Pocock \cite{pocock}, Peto \cite{peto}, and O'Brien-Fleming \cite{obf} spending functions. Several sequential and adaptive Bayesian approaches also exist \cite[e.g.][]{tan02, thall_wathen05, thall_wathen07, thall_wathen08}, though many Bayesian designed studies will ultimately use thresholds corresponding to frequentist characteristics, making the results of the experiment difficult to interpret \cite{valen08}. 
\\

Among sequential tests that do not require a sample size to be calculated \emph{a priori}, perhaps the most well-known is Wald's sequential probability ratio test (SPRT) \cite{wald_sprt}, which allows for continuous monitoring of a test between two pre-specified hypotheses, with the experiment terminating when the test statistic crosses one of two thresholds - one corresponding to accepting the null hypothesis, and the other the alternative. The test, however, only considers two simple hypotheses, and is therefore challenging to apply in the more commonplace scenarios of composite hypotheses presented above, for example. Several extensions to the SPRT have been developed \cite[e.g.][]{maxsprt, hoel76, meeker81}, and many of these do extend the SPRT to allow for one-sided hypothesis testing. An alternative, but very similar approach was developed by Cheng and Shen \cite{chengshen} under a Bayesian decision-theoretic framework, which is very appealing for the problem presented here, and which I discuss in more detail below. 

\subsection*{Sequential Monitoring in A/B Testing}

While fixed-sample designs have historically been the most popular approach, sequential designs have been around since the first half of last century - for roughly 70 years dating back to Wald's SPRT \cite{wald_sprt} in 1945, and there are several other examples of this type of methodology and/or extensions of the original SPRT in the literature \citep[see, e.g.,][to name a few]{wald_sprt, maxsprt, hoel76, meeker81, armitage58, armitage_book, optimizely}. The appeal of sequential sample size designs is that, on average, they allow researchers to answer questions with smaller sample sizes than do fixed sample size designs \cite[Ch. 87]{trials_handbook}. The reason, however, that fixed sample designs have been most popular is because sequential sample size designs are limited to settings where outcomes can be categorized as successes or failures, and where that categorization can be made shortly after enrollment in the study. The categorization is important, because continuing the experiment to include the next person or group of persons is dependent on being able to observe the outcome of all people that have previously participated in the study \cite[Ch. 87]{trials_handbook}. Sequential designs are only feasible when there is a large group of people who are suitable to be participants in the study in short order following the observed outcomes of those who have already participated in the study. Additionally, each study participant's outcome must be observable in a short amount of time following entry to the study, and the treatments (or variants) administered to the exposed group(s) can't extend over a long period of time because the length of the study would be increased dramatically, which would be quite impractical \cite[Ch. 87]{trials_handbook}. Further, if a researcher was also implementing an adaptive randomization scheme on top of the sequential design, a study taking place over too long a period of time could become subject to drift which could cause the adaptive randomization to perform poorly and bias results \cite{thall_wathen07}.
\\

A great deal of literature on study design and hypothesis testing can be found not just in purely statistical journals, but also biostatistical and medical journals. This is attributable to the development of clinical trial methodology in the medical field; after all, what is a clinical trial but a (possibly multivariate) A/B test comparing the performance of one or more drugs for treating a particular disease to a placebo or other type of control group. This, then, explains why so many experiments take on a fixed-sample design. It can be very challenging, both administratively and analytically, to conduct a sequential or adaptive clinical trial because of either patient recruitment/enrollment issues, or if the outcome of interest isn't immediately available after administering a treatment to a patient, or if administering the treatment is a process that takes several weeks. That said, there is a sizable amount of research on sequential and adaptive designs in clinical trials (most of the works cited to this point, for example), and interest is continually growing as researchers are constantly looking for better ways to more quickly and ethically identify the most helpful treatments for numerous life-threatening diseases and conditions. Indeed, a PubMed search shows a steady increase in the number of publications containing the phrases ``sequential design'' and ``adaptive design'' over the past 25+ years \cite{pubmed_trend}.
\\

Therefore, when looking to develop a method for a multivariate A/B testing procedure, it was natural to look to the medical and clinical trials literature. In particular, I opted to base the methodology on the work of Cheng and Shen \cite{chengshen} by generalizing their method to accommodate a multivariate A/B test.

\subsection*{A Bayesian Sequential Design for A/B Testing}

The title of this section plays on the title of Cheng's and Shen's 2005 paper, Bayesian Adaptive Designs for Clinical Trials \cite{chengshen}. Despite their use of the word ``adaptive'' in the title, they do not develop nor implement adaptive procedures in the way defined here. Rather, they develop a Bayesian theoretic approach to a sequential two-arm clinical trial, or A/B test, in which the total sample size in the experiment is not fixed \emph{a priori}, nor is a maximum sample size imposed. Their method is as follows \cite{chengshen}.
\\

Consider a one-sided hypothesis test similar to those defined earlier in which the researcher wishes to test:

$$H_{0}: \mu_{A} + \theta_{0} \geq \mu_{B} \hspace{0.75cm}\mathrm{vs.}\hspace{0.75cm} H_{1}: \mu_{A} < \mu_{B}.$$

Define $\theta = \mu_{B} - \mu_{A}$, where, as before, $\mu_{B}$ represents the average response rate in the exposed group, B; $\mu_{A}$ represents the average response rate in the control group, A; and $\theta_{0}$ is some constant $\geq 0$. For simplicity let $\theta_{0} = 0$ throughout, so that the hypothesis test can be rewritten as:

%$$H_{0}: \theta \leq \theta_{0} \hspace{0.75cm}\mathrm{vs.}\hspace{0.75cm} H_{1}: \theta > 0,$$
%
%where $\theta_{0}$ is some constant $\geq 0$. These hypotheses include two scenarios: with ($\theta_{0} > 0$) and without ($\theta_{0} = 0$) the range of equivalence. For simplicity we assume $\theta_{0} = 0$, so that we can rewrite the hypotheses one last time as:

$$H_{0}: \theta \leq 0 \hspace{0.75cm}\mathrm{vs.}\hspace{0.75cm} H_{1}: \theta > 0.$$

Let $A$ and $R$ denote the actions of either accepting or rejecting the null hypothesis, respectively. The authors then define the following loss functions:

\begin{equation*}
	L(\theta, A) = \begin{cases}
		0, & \text{if $\theta \leq 0$};\\
		K_{1}, & \text{if $\theta > 0$};
	\end{cases}
\end{equation*}
\vspace{0.25cm}
\begin{equation*}
	L(\theta, R) = \begin{cases}
		K_{0}, & \text{if $\theta \leq 0$};\\
		0, & \text{if $\theta > 0$};
	\end{cases}
\end{equation*}

where $K_{0}$ and $K_{1}$ are the losses for making a Type I and Type II error, respectively.
\\

The goal is to devise a stopping rule to minimize the loss. Define $\mathcal{X}_{j} = \{X_{1}, \ldots, X_{j}\}$ as the cumulative data collected up to step $j$ of the experiment (in other words, after $j$ looks at the data), and define the corresponding information set at that time as the $\sigma$-algebra $\mathcal{F}_{j} = \sigma(\mathcal{X}_{j})$. The total cost (or loss) of stopping the experiment after the $j$-th look at the data is the cost of enrolling everyone in the study up to and including that point, plus the loss of either accepting or rejecting the the null hypothesis at that point, whichever is smaller:

\begin{equation}\label{eq:lstop}
L_{stop}(\mathcal{X}_{j}) = 2K_{2}\sum_{i=1}^{j}B_{i} + \mathrm{min}\big[E\{L(\theta, A) | \mathcal{F}_{j}\}, E\{L(\theta, R) | \mathcal{F}_{j}\}\big],
\end{equation}

 where $K_{2}$ is the cost of enrolling one person into the study, and $B_{i}$ is the number of additional people involved in the study between looks $i-1$ and $i$ of the experiment. Further, note that
 
 $$E\{L(\theta, A) | \mathcal{F}_{j}\} = K_{1}P(\theta > 0 | \mathcal{F}_{j}) \hspace{0.5cm}\text{and}\hspace{0.5cm} E\{L(\theta, R) | \mathcal{F}_{j}\} = K_{0}P(\theta \leq 0 | \mathcal{F}_{j}).$$

The loss of continuing the experiment to allow for a $(j+1)$-th look at the data can be expressed as:

\begin{equation}\label{eq:lcont}
L_{cont}(\mathcal{X}_{j}) = 2K_{2}\sum_{i=1}^{j+1}B_{i} + E\Big(\mathrm{min}\big[E\{L(\theta, A) | \mathcal{F}_{j+1}\}, E\{L(\theta, R) | \mathcal{F}_{j+1}\}\big] \Big| \mathcal{F}_{j} \Big),
\end{equation}

where the outside expectation in \eqref{eq:lcont} is taken with respect to the posterior predictive distribution of $\theta$.
\\

The decision to reject or accept the null hypothesis at the $j$-th look would proceed as follows: If the expected loss of accepting the null hypothesis at step $j$ is less than or equal to the expected loss of rejecting the null hypothesis at step $j$, then accept the null; otherwise, reject. Notationally, the rejection region, $R_{j}$, can be defined as:

\begin{equation}\label{eq:rejreg}
R_{j} = \bigg\{\mathcal{X}_{j} : \frac{P(\theta > 0 | \mathcal{F}_{j})}{P(\theta \leq 0 | \mathcal{F}_{j})} \geq \frac{K_{0}}{K_{1}}\bigg\}, \hspace{0.25cm} j = 1, 2, \ldots
\end{equation}
\\
The authors show that the Type I and Type II error rates can be controlled by judicious selection of the parameters $(K_{0}, K_{1}, K_{2})$. Specifically, they show that for a desired Type I error rate of $\alpha$, setting the ratio $K_{0}/K_{1} = (1-\alpha)/\alpha$ will place an upper bound on the Type I error of their test at $\alpha$, while a grid search over the value of the ratio $K_{2}/K_{1}$ can be performed to obtain the desired Type II error rate.
\\

To select the optimal study design that minimizes the expected loss, the following two-step procedure is followed:

\begin{enumerate}

\item If $L_{stop}(\mathcal{X}_{j}) \leq L_{cont}(\mathcal{X}_{j})$, the experiment is stopped at interim look $j$. If at that point $\mathcal{X}_{j}$ is in the rejection region, $R_{j}$, reject the null; otherwise accept.

\item If $L_{stop}(\mathcal{X}_{j}) > L_{cont}(\mathcal{X}_{j})$, then continue the experiment for one more look at the data and repeat steps 1 and 2.

\end{enumerate}

Because of the iterative nature of this study design, the stopping time of the trial, denoted $M$, is not pre-fixed but rather is a random integer. The authors show in their Theorem 1 that for $K_{2} > 0$, $M$ is a stopping time satisfying $P(M < \infty \hspace{0.1cm}| \hspace{0.1cm} \theta) = 1$. In other words, if $K_{2} > 0$, then the experiment is guaranteed to terminate at some point and not go on forever.

\subsubsection*{Extensions}

In this section I discuss my adaptation of the Cheng and Shen methodology outlined above. First, note the form of the rejection region, $R_{j}$, defined in \eqref{eq:rejreg}. The test statistic checked at every interim look is:

\begin{equation}\label{eq:bf1}
\frac{P(\theta > 0 | \mathcal{F}_{j})}{P(\theta \leq 0 | \mathcal{F}_{j})},
\end{equation}

which is easily identified as the Bayes factor for a model with $\theta > 0$ compared to one with $\theta \leq 0$ under an assumption of equal prior odds for both hypotheses, a reasonable and uninformative \emph{a priori} assumption with respect to favoring one hypothesis over the other. Note, also, the striking similarity between this test statistic and that of the classic Wald SPRT \cite{wald_sprt} which can also be interpreted as a Bayes factor, although comparing two simple hypotheses. It should not be surprising, then, that the critical values for those two test statistics are so similar. For Wald's SPRT, the null hypothesis is rejected once the Bayes factor is greater than $(1-\beta)/\alpha$, and is accepted if the Bayes factor is less than $\beta/(1-\alpha)$ where $\alpha$ and $\beta$ are the Type I and Type II errors, respectively. If at any given interim check of the data either of those thresholds is crossed, the experiment is stopped; otherwise more subjects are enrolled until a decision can be made.
\\

Second, I examine the stopping rule derived by Cheng and Shen; specifically that the study is stopped once $L_{stop}(\mathcal{X}_{j}) \leq L_{cont}(\mathcal{X}_{j})$, or equivalently, when

\begin{equation}\label{eq:stop1}
\begin{split}
& 2K_{2}B_{j+1} + E\Big(\mathrm{min}\big[E\{L(\theta, A) | \mathcal{F}_{j+1}\}, E\{L(\theta, R) | \mathcal{F}_{j+1}\}\big] \Big| \mathcal{F}_{j} \Big) \\
\vspace{0.5cm}
& \geq \hspace{0.1cm}  \mathrm{min}\big[E\{L(\theta, A) | \mathcal{F}_{j}\}, E\{L(\theta, R) | \mathcal{F}_{j}\}\big].
 \end{split}
\end{equation}

In letting: 

$$Y_{j} = \mathrm{min}\big[E\{L(\theta, A) | \mathcal{F}_{j}\}, E\{L(\theta, R) | \mathcal{F}_{j}\}\big],$$ 

the authors show, by Jensen's inequality, that:

\begin{equation}\label{eq:jensen}
E(Y_{j+1} | \mathcal{F}_{j}) \leq \mathrm{min}\Big(E\big[E\{L(\theta, A) | \mathcal{F}_{j+1}\} \big| \mathcal{F}_{j}\big], E\big[E\{L(\theta, R) | \mathcal{F}_{j+1}\} \big| \mathcal{F}_{j} \big] \Big) = Y_{j}.
\end{equation}

Following the notation in \eqref{eq:jensen}, the stopping rule \eqref{eq:stop1} can be rewritten as

\begin{equation}
2K_{2}B_{j+1} \geq Y_{j} - E(Y_{j+1} | \mathcal{F}_{j}).
\end{equation}

By the proof of Theorem 1 in \cite{chengshen} (or \eqref{eq:jensen} above), $0 \leq Y_{j} - E(Y_{j+1} | \mathcal{F}_{j})$, with equality coming by taking the limit as $j \rightarrow \infty$ \cite{chengshen}. By taking $K_{2} > 0$ it can be ensured that the experiment will reach a stopping time, $M < \infty$. However, I argue that, in the setting of digital marketing, it is not unreasonable to set $K_{2} = 0$. 
\\

First, consider that the primary concern here is that of a mobile or digital marketer serving ads to his customers. If that marketer has the ability to send an ad to as many as 500,000 customers, the cost of sending a campaign out to 200,000 people versus 100,000, for example, is practically negligible; even if $K_{2}$ is not truly zero, it should be very, very close. Secondly, a marketer will likely be operating from the standpoint of wanting to run a campaign for some number of pre-specified weeks, but with the option to stop if results show themselves to be significant before that point. In practice, therefore, the experiment would never run for an infinite amount of time. The sequential monitoring procedure simply allows the marketer to monitor the campaign continuously and stop the experiment at any time, with all of the results maintaining statistical validity with respect to the Type I error rate whenever the campaign is halted.
\\

Further, by recognizing that the test statistic \eqref{eq:bf1} can be interpreted as a Bayes factor, I propose that the test can easily be extended to two-sided hypotheses of the form:

$$H_{0}: \theta = 0 \hspace{0.75cm}\mathrm{vs.}\hspace{0.75cm} H_{1}: \theta \neq 0,$$

by altering the test statistic to:

\begin{equation}\label{eq:bf2}
\frac{P(\theta \neq 0 | \mathcal{F}_{j})}{P(\theta = 0 | \mathcal{F}_{j})},
\end{equation}

with a rejection region of 

\begin{equation}\label{eq:rejreg2}
R_{j} = \bigg\{\mathcal{X}_{j} : \frac{P(\theta \neq 0 | \mathcal{F}_{j})}{P(\theta = 0 | \mathcal{F}_{j})} \geq c_{\alpha/2}\bigg\}, \hspace{0.25cm} j = 1, 2, \ldots
\end{equation}
\\
where $c_{\alpha/2}$ is a threshold cutoff, the value of which is selected to ensure a desired Type I error rate, $\alpha$. Because I set $K_{2} = 0$, the rejection region above also doubles as the stopping rule: If $\mathcal{X}_{j} \in R_{j}$, stop the experiment and reject the null hypothesis; otherwise, continue enrolling customers into the study. Such a stopping rule is very similar to that of the Wald SPRT. Via simulation (results in later section), I show that the same $c_{\alpha/2} = (1-\alpha/2)/(\alpha/2)$ stopping rule adopted by Cheng and Shen maintains the desired Type I error rate at $\alpha$. Additionally, because the hypothesis test is two-sided, the stopping rule encompasses both superiority and inferiority of the treatment arm.

\subsubsection*{Estimation}

Any number of estimation procedures would work under this framework. I prefer a Bayesian approach to the problem, because \emph{a posteriori} comparisons and transformations are greatly simplified under a Bayesian framework, and because it is also appealing to assign a probability of being the best performing arm to each arm of a multivariate A/B test - something that is not feasible in a frequentist framework. Specifically, I choose to estimate parameters via a logistic regression, which provides the flexibility to additionally control for any demographic variables available if desired, as well as to extend the model to easily include multiple treatment groups and any interactions between treatment group and demographics, all in one single estimation procedure. The simple two-arm model can take the following form:

\begin{equation}\label{eq:logit1}
\mathrm{logit}(p_{i}) = \alpha + \beta*\mathrm{trt}_{i} + \sum_{j=1}^{k}\gamma_{j}z_{j},
\end{equation}

where $p_{i}$ is the probability of a success (e.g., a conversion), $\mathrm{trt}_{i}$ is a variable capturing whether person $i$ is in the treatment or control group, each $z_{j}$ is one of $k$ optional demographic variables to additionally control for, and $\Theta = (\alpha, \beta, \gamma_{1}, \ldots, \gamma_{k})$ is a vector of parameters to be estimated by the model. The main parameter of interest is $\beta$, and the hypothesis test can be written:

$$H_{0}: \beta = 0 \hspace{0.75cm}\mathrm{vs.}\hspace{0.75cm} H_{1}: \beta \neq 0.$$

Note that $\beta$ captures the log-odds of success for the treatment group compared to the control group, though it can easily be transformed to give the estimated probability of success in each group. The Bayes factor \eqref{eq:bf2} can be calculated by comparing the model that includes the treatment variable to one that does not. As model complexity grows, and as the model is extended to accommodate multivariate A/B tests, a fully Bayesian framework can prove computationally intensive, particularly when evaluating tens or hundreds of campaigns at once. In practice, therefore, the parameters of the model can be estimated using standard statistical software, e.g., via the \texttt{glm} function in R, with samples from the posterior distribution of the parameters generated using the functionality of R's \texttt{arm} package. Additionally, the Bayes factor can be approximated by using Schwarz's Bayesian Information Criterion (BIC) from the fitted models \cite{bic78, kass93, raftery99}, which also greatly reduces computation time.

\subsection*{Multivariate A/B Tests and Multiple Testing}

The methodology outlined to this point covers only a simple two-arm A/B test. In this section, I present extensions of the results above to multivariate ($> 2$ arm) A/B tests.
\\

The main multiple comparisons problem is that the probability an experimenter incorrectly concludes that there is at least one statistically significant result across a group of tests, even when there are in fact no real associations, increases with each additional test \cite{mehta, armitage69, gelman12}. A related issue is that, in a setting where actual non-zero effects do exist for some of the tests, an experimenter applying multiple tests may find additional statistically significant results that are not real \cite{gelman12}.
\\

Several approaches have been developed to address the multiple comparisons issue. The most common of these is the Bonferroni correction, which divides the desired Type I error rate, $\alpha$, by the number of hypotheses being tested, $m$, yielding an adjusted Type I error rate of $\alpha^{*} = \alpha/m$ for each of the $m$ tests. Other methods for controlling the Type I error rate include the \v{S}id\'{a}k correction and the Holm-Bonferroni correction, among others \cite[see, e.g.,][]{hsu_multcomp}. All of these procedures aim to control the familywise error rate (FWER) across the group of tests at $\alpha$. The main drawback to Bonferroni-type procedures is that the Type I error problem is adjusted at the expense of the Type II error, meaning the power of the statistical test can be greatly reduced using one of these adjustments. An alternative approach to the multiple comparisons problem is to, instead of controlling the overall Type I error rate, control the False Discovery Rate (FDR). In doing so, the researcher is able to ensure that among all tests declared statistically significant, only a certain percentage are expected to be false positives \cite{benj_hoch95}. Under FDR adjustment, the test is less conservative with respect to Type I error, though the researcher will have confidence that a high percentage of the significant results detected represent real differences. A common thread between methods controlling the FWER and those controlling the FDR is that both do so by means of a p-value adjustment.
\\

An alternative approach to the multiple comparisons problem is described by Gelman, et. al. \cite{gelman12}, wherein the authors shift the multiple comparisons problem into a modeling issue and address it by operating from a Bayesian multilevel modeling framework. Unlike FWER or FDR adjustments which, in essence, keep point estimates stationary and adjust for multiple comparisons by making confidence intervals wider (or, equivalently, by adjusting the p-values), the multilevel modeling approach instead shifts point estimates and their corresponding intervals closer together - a process known as ``shrinkage'' or ``partial pooling.'' In this way, the estimates from a multilevel model naturally make comparisons more conservative by increasing the likelihood that any intervals for comparisons contain zero \cite{gelman12}. At the same time, the method does not correct for multiple comparisons at the expense of the power of the test, unlike many of the traditional methods discussed above \cite{gelman12}.
\\

Rather than operating from the standard Type I error paradigm, Gelman, et. al. argue that more troublesome issues are when either: 1. estimators have the incorrect sign, or 2. estimators declare an effect to be small when it is actually large, or declare it is actually large when in fact it is near zero. They dub these errors of sign, and errors of magnitude - Type S and Type M errors, respectively. While the multilevel Bayesian approach addresses Type S and Type M errors, it also does well to control for the conventional Type I error while maintaining good power. Additionally, given the Bayesian nature of the analysis proposed thus far, I find it desirable to address the multiple comparison issue in a very natural Bayesian approach as opposed to making p-value adjustments since p-values are inherently very non-Bayesian. Simulation results summarizing the performance of the full multivariate A/B testing methodology are presented in a later section, but first consider how the approach of Gelman, et. al. \cite{gelman12} can be applied to the method presented thus far.
\\

Recall that in a simple two-arm A/B test, the comparison between the treatment and control group can be made by fitting the model given by \eqref{eq:logit1}. In an $(m+1)$-arm A/B test $(m > 1)$, a natural extension to \eqref{eq:logit1} would be:

\begin{equation}\label{eq:logitm}
\mathrm{logit}(p_{i}) = \alpha + \sum_{r=1}^{m}\beta_{r}*\mathrm{trt}_{i,r} + \sum_{j=1}^{k}\gamma_{j}z_{j},
\end{equation}

where now $\mathrm{trt}_{i,r}$ is an indicator for whether person $i$ was on arm $\mathrm{B}_{r}$, and $\beta_{r}$ is the corresponding coefficient capturing the log-odds of success in group $\mathrm{B}_{r}$ compared to the baseline group, A. However to account for multiple comparisons under \eqref{eq:logitm}, some kind of p-value adjustment would need to be made to adjust the FWER or FDR of the test. Instead, I propose the following multilevel model based on \cite{gelman12}:

\begin{equation}\label{eq:logitpool}
\mathrm{logit}(p_{i}) = \alpha_{r[i]} + \sum_{j=1}^{k}\gamma_{j}z_{j},
\end{equation}

where $\alpha_{r[i]}$ captures the log-odds of success in person $i$'s arm of the experiment. And 

$$\alpha_{r} \sim N(\mu, \sigma_{\alpha}^{2}).$$

The caveat is that now $r = 1, 2, \ldots, m, m+1$, meaning that the control group's success rate is also part of the pooling estimation. I believe this is reasonable in the sense that it is not unrealistic to assume these are all realizations from a common distribution - it is the same outcome being measured in each arm of the experiment, and it seems plausible that, across multiple experiments, some arms might perform better than baseline, some worse, and some very similarly. Assume $\alpha_{1}$ captures the log-odds in the control group, then the multiple hypotheses being tested are of the form:

\begin{equation}\label{eq:pooledtests}
H_{0}: \alpha_{r}-\alpha_{1} = 0 \hspace{0.75cm}\mathrm{vs.}\hspace{0.75cm} H_{1}: \alpha_{r}-\alpha_{1} \neq 0,\hspace{0.75cm} \forall r \geq 2
\end{equation}

As before, the model \eqref{eq:logitpool} can be fit in R, in this case by using the \texttt{glmer} function in the \texttt{lme4} package, with posterior simulations given by the \texttt{arm} package. It is also easy to extend the model to obtain subgroup estimates by treatment group. For more details on the multiple testing procedure, the reader is encouraged to see \cite{gelman12}. 

\section*{Simulation Results}

In this section simulation results for the full method outlined above are presented, combining the estimation procedure of \eqref{eq:logitpool} with the decision rule given by \eqref{eq:rejreg2}. Table \ref{table1} summarizes the average expected outcomes of five hypothetical multivariate A/B experiments, each of which is a five-arm (four treatment + one control) experiment with equal allocation in each arm. The results presented for each multivariate A/B experiment represent an average over 1,000 simulated trials. The hypotheses being tested are two-sided hypothesis tests taking the form of \eqref{eq:pooledtests}. All hypothesis tests are carried out at the $\alpha = 0.05$ level, yielding a cutoff value of $c_{\alpha/2}=39$. For convenience a maximum of 20,000 people in each arm is enforced, and the data is checked after every 500 enrollments in each arm, for a total of 40 interim looks at the data. The fields in Table \ref{table1} are as follows:
\\

\begin{table}[h]
\small
\centering
\caption{Results of Interest from Simulated Experiments}
\label{defs}
\hspace*{-0.5cm}\begin{tabular}{|l | l|}
\hline
Field                      & Description                                                                                                          \\
\hline
$p_{0}$                              & \begin{tabular}[c]{@{}l@{}}Probability of success in the control group, i.e. baseline \\ success rate.\end{tabular}                                                      \\
\hline
$p_{r}$                              & Probability of success in treatment group $r$.                                                                                                                           \\
\hline
%P(reject)                            & \begin{tabular}[c]{@{}l@{}}Probability of rejecting the null hypothesis in favor of the \\ alternative for each arm.\end{tabular}                                        \\
%\hline
Power                                & \begin{tabular}[c]{@{}l@{}}Estimated power to detect a difference between each arm and \\ baseline; formally P(reject $H_{0} \hspace{0.1cm}|\hspace{0.1cm} H_{1}$ True).\end{tabular}              \\
\hline
Type I Error $(\hat{\alpha})$                         & \begin{tabular}[c]{@{}l@{}}Estimated Type I Error for each arm; \\ formally P(reject $H_{0} \hspace{0.1cm}|\hspace{0.1cm} H_{0}$ True)\end{tabular}                                                \\
\hline
$\overline{N}$                       & Average sample size required in each arm across all simulated trials                                                                                                     \\
\hline
Fixed-Sample Avg. Power              & \begin{tabular}[c]{@{}l@{}}Average power for a fixed-sample design given the observed $N$ \\ from each simulated trial without multiple testing correction.\end{tabular} \\
\hline
Fixed-Sample Avg. Power (Bonferroni) & \begin{tabular}[c]{@{}l@{}}Average power for a fixed-sample design given the observed $N$ \\ from each simulated trial with a Bonferroni correction.\end{tabular}        \\
\hline
FWER                                 & \begin{tabular}[c]{@{}l@{}}Familywise Error Rate. Percentage of simulated trials producing \\ at least one false positive.\end{tabular}                                \\
\hline
FDR                                  & \begin{tabular}[c]{@{}l@{}}False Discovery Rate. Percentage of statistically significant results \\ that are actually false positives.\end{tabular}                      \\
\hline
Per-test $\hat{\alpha}$              & Average Type I Error rate across all arms where $H_{0}$ is true.\\
\hline                                                                                                                     
\end{tabular}
\end{table}

\newpage

\begin{landscape}
\begin{table}[ht]
\small
\centering
\vspace*{-1cm}
\caption{Five-arm Simulation Results}
\label{table1}
\hspace*{-1cm}\begin{tabular}{ccccrccccc}
\multicolumn{1}{c}{$p_{0}$} & \multicolumn{1}{c}{$p_{r}$} & \multicolumn{1}{c}{\begin{tabular}{c}Power\\ (1-$\hat{\beta}$)\end{tabular}} & \multicolumn{1}{c}{\begin{tabular}{c}Type I Error\\ ($\hat{\alpha}$)\end{tabular}} & \multicolumn{1}{c}{$\overline{N}$} & \multicolumn{1}{c}{\begin{tabular}{c}Fixed-Sample\\ Avg. Power\end{tabular}} & \multicolumn{1}{c}{\begin{tabular}{c}Fixed-Sample\\ Avg.Power\\ (Bonferroni)\end{tabular}} & \multicolumn{1}{c}{FWER} & \multicolumn{1}{c}{FDR} & \multicolumn{1}{c}{Per-test $\hat{\alpha}$} \\
\hline
\\
\multirow{4}{*}{0.50}       & 0.48                                    & 0.820                                                                                 & --                                                                                         & 12,396.0                           & 0.805                                                                                 & 0.673                                                                                                      & \multirow{4}{*}{0.024}   & \multirow{4}{*}{0.013}  & \multirow{4}{*}{0.012}                      \\
                            & 0.50                                       & --                                                                                   & 0.014                                                                                       & 19,817.5                           & --                                                                                   & --                                                                                                        &                          &                         &                                             \\
                            & 0.50                                           & --                                                                                   & 0.010                                                                                       & 19,875.0                           & --                                                                                   & --                                                                                                        &                          &                         &                                             \\
                            & 0.53                                     & 1.000                                                                                 & --                                                                                         & 6,542.5                            & 0.849                                                                                 & 0.731                                                                                                      &                          &                         &                                             \\[4ex]
\multirow{4}{*}{0.250}      & 0.235                               & 0.617                                                                                 & --                                                                                         & 14,816.5                           & 0.796                                                                                 & 0.654                                                                                                      & \multirow{4}{*}{0.011}   & \multirow{4}{*}{0.004}  & \multirow{4}{*}{0.011}                      \\
                            & 0.250                                          & --                                                                                   & 0.011                                                                                       & 19,858.0                           & --                                                                                   & --                                                                                                        &                          &                         &                                             \\
                            & 0.270                                          & 0.935                                                                                 & --                                                                                         & 9,522.0                            & 0.784                                                                                 & 0.650                                                                                                      &                          &                         &                                             \\
                            & 0.280                                  & 1.000                                                                                 & --                                                                                         & 4,953.5                            & 0.843                                                                                 & 0.723                                                                                                      &                          &                         &                                             \\[4ex]
\multirow{4}{*}{0.1000}     & 0.0875                             & 0.864                                                                                 & --                                                                                         & 11,638.0                           & 0.816                                                                                 & 0.694                                                                                                      & \multirow{4}{*}{0.008}   & \multirow{4}{*}{0.003}  & \multirow{4}{*}{0.008}                      \\
                            & 0.1000                             & --                                                                                   & 0.008                                                                                       & 19,898.0                           & --                                                                                   & --                                                                                                        &                          &                         &                                             \\
                            & 0.1175                                 & 0.994                                                                                 & --                                                                                         & 6,513.0                            & 0.798                                                                                 & 0.666                                                                                                      &                          &                         &                                             \\
                            & 0.1250                              & 1.000                                                                                 & --                                                                                         & 3,716.5                            & 0.846                                                                                 & 0.728                                                                                                      &                          &                         &                                             \\[4ex]
\multirow{4}{*}{0.0500}     & 0.0425                             & 0.658                                                                                 & --                                                                                         & 14,336.5                           & 0.793                                                                                 & 0.654                                                                                                      & \multirow{4}{*}{0.013}   & \multirow{4}{*}{0.005}  & \multirow{4}{*}{0.013}                      \\
                            & 0.0500                                  & --                                                                                   & 0.013                                                                                       & 19,874.5                           & --                                                                                   & --                                                                                                        &                          &                         &                                             \\
                            & 0.0600                                    & 0.910                                                                                 & --                                                                                         & 10,315.5                           & 0.786                                                                                 & 0.654                                                                                                      &                          &                         &                                             \\
                            & 0.0650                        & 1.000                                                                                 & --                                                                                         & 5,548.0                            & 0.837                                                                                 & 0.717                                                                                                      &                          &                         &                                             \\[4ex]
\multirow{4}{*}{0.0250}     & 0.0250                          & --                                                                                   & 0.003                                                                                       & 19,975.0                           & --                                                                                   & --                                                                                                        & \multirow{4}{*}{0.003}   & \multirow{4}{*}{0.001}  & \multirow{4}{*}{0.003}                      \\
                            & 0.0325                  & 0.929                                                                                 & --                                                                                         & 10,015.5                           & 0.791                                                                                 & 0.659                                                                                                      &                          &                         &                                             \\
                            & 0.0400                            & 1.000                                                                                 & --                                                                                         & 3,382.5                            & 0.864                                                                                 & 0.751                                                                                                      &                          &                         &                                             \\
                            & 0.0435                          & 1.000                                                                                 & --                                                                                         & 2,612.5                            & 0.899                                                                                 & 0.803                                                                                                      &                          &                         &                                             \\
\end{tabular}%\hspace*{-1cm}
\end{table}
\end{landscape}

From Table \ref{table1} it's clear that the proposed method almost universally outperforms a standard fixed-sample design under a Bonferonni correction with respect to power. Surprisingly, even though the proposed method accounts for multiple testing, it also outperforms a standard fixed-sample design with no multiple testing adjustment with respect to power and Type I error in all but two of the 20 arms. The FDR rates are also quite good across all experiments, although it is important to note that only the test with $p_{0} = 0.50$ had more than one arm where the null hypothesis was true. Still, the results in Table \ref{table1} provide a good overview of the performance of the methodology where there are only a handful of hypotheses being simultaneously tested (only 4 comparisons being made) and where the null hypothesis is not true in most arms.
\\

Table \ref{table3} below displays analogous results to those in Table \ref{table1} above, but for 11-arm (ten treatment + one control) experiments with equal allocation in each arm. In each experiment, the null hypothesis of no treatment effect is true in five out of ten of the arms. To conserve space, the results for only two (instead of five) such experiments are presented: one with $p_{0} = 0.5$, and the other with $p_{0} = 0.05$. As before, the results presented for each multivariate A/B experiment represent an average over 1,000 simulated trials; all hypothesis tests were two-sided and carried out at an $\alpha = 0.05$ level; and a maximum of 20,000 people in each arm was enforced, with interim checks at every 500 enrollments for each arm. Even in this scenario, the proposed method performs no worse than a Bonferroni-corrected experiment with respect to power, and for ``larger'' differences between treatment and control the proposed method far outperforms it - even doing better, again, than fixed-sample designs with no multiple testing correction. Of particular note, the FWER is controlled below the $\alpha=0.05$ level, and the FDR is quite low, between 1\% and 1.5\% for both experiments. 
\\

Another attractive aspect of this method is that in fitting the logistic model under a Bayesian framework, it is easy to obtain, via posterior simulation, the probability corresponding to each arm that it is the best performing arm in the study at each interim check, or $P(p_{r^{*}} = \max_{r}(p_{r}) \hspace{0.1cm} | \hspace{0.1cm} \mathcal{F}_{j})$.\comment{This can be done at every interim check of the data, as the investigator monitors the experiment to determine if it can/should be terminated.} The result is a procedure that not only quantifies lift over baseline and whether that lift is statistically significant or not, but also through these posterior probabilities provides a means to compare the strength of evidence against the null hypothesis for each of the arms in the study to one another. In other words, if two arms show significant lift over baseline and one has a point estimate of 0.03 and the other 0.02, without digging much deeper it would be reasonable to conclude that they performed very similarly to each other. If however, you have readily available the fact that $P(p_{0.03} = \max_{r}(p_{r})\hspace{0.1cm} | \hspace{0.1cm} \mathcal{F}_{j})=0.70$, while $P(p_{0.02} = \max_{r}(p_{r})\hspace{0.1cm} | \hspace{0.1cm} \mathcal{F}_{j})=0.25$, the arm with the point estimate of 0.03 now carries more weight than if one only knew the point estimates. Figure \ref{pbest_plot} below shows how these probabilities change throughout a trial as the data collected grows. Specifically, Figure \ref{pbest_plot} plots the average of $P(p_{r^{*}} = \max_{r}(p_{r})\hspace{0.1cm} | \hspace{0.1cm} \mathcal{F}_{j})$ for each $j$ and for each treatment arm corresponding to the simulated trial in Table \ref{table3} with $p_{0}=0.5$.

\newpage

\begin{landscape}
\begin{table}[ht]
\small
\centering
\vspace*{0.15cm}
\caption{11-arm Simulation Results}
\label{table3}
\hspace*{-1.5cm}\begin{tabular}{ccccrccccc}
\multicolumn{1}{c}{$p_{0}$} & \multicolumn{1}{c}{$p_{r}$}  & \multicolumn{1}{c}{\begin{tabular}{c}Power\\ (1-$\hat{\beta}$)\end{tabular}} & \multicolumn{1}{c}{\begin{tabular}{c}Type I Error\\ ($\hat{\alpha}$)\end{tabular}} & \multicolumn{1}{c}{$\overline{N}$} & \multicolumn{1}{c}{\begin{tabular}{c}Fixed-Sample\\ Avg. Power\end{tabular}} & \multicolumn{1}{c}{\begin{tabular}{c}Fixed-Sample\\ Avg.Power\\ (Bonferroni)\end{tabular}} & \multicolumn{1}{c}{FWER} & \multicolumn{1}{c}{FDR} & \multicolumn{1}{c}{Per-test $\hat{\alpha}$} \\
\hline
\\
\multirow{10}{*}{0.50}      & 0.48                               & 0.820                                                                                 & --                                                                                          & 12,668.5                           & 0.816                                                                                 & 0.598                                                                                                      & \multirow{10}{*}{0.038}  & \multirow{10}{*}{0.014} & \multirow{10}{*}{0.009}                     \\
                            & 0.49                                          & 0.179                                                                                 & --                                                                                          & 18,576.5                           & 0.484                                                                                 & 0.192                                                                                                      &                          &                         &                                             \\
                            & 0.50                                       & --                                                                                    & 0.011                                                                                       & 19,914.0                           & --                                                                                    & --                                                                                                         &                          &                         &                                             \\
                            & 0.50                              & --                                                                                    & 0.006                                                                                       & 19,907.5                           & --                                                                                    & --                                                                                                         &                          &                         &                                             \\
                            & 0.50                                      & --                                                                                    & 0.005                                                                                       & 19,946.0                           & --                                                                                    & --                                                                                                         &                          &                         &                                             \\
                            & 0.50                                      & --                                                                                    & 0.009                                                                                       & 19,888.5                           & --                                                                                    & --                                                                                                         &                          &                         &                                             \\
                            & 0.50                                           & --                                                                                    & 0.013                                                                                       & 19,864.0                           & --                                                                                    & --                                                                                                         &                          &                         &                                             \\
                            & 0.51                                         & 0.202                                                                                 & --                                                                                          & 18,390.0                           & 0.480                                                                                 & 0.190                                                                                                      &                          &                         &                                             \\
                            & 0.52                                        & 0.844                                                                                 & --                                                                                          & 11,983.0                           & 0.793                                                                                 & 0.565                                                                                                      &                          &                         &                                             \\
                            & 0.53                                         & 0.999                                                                                 & --                                                                                          & 6,802.5                            & 0.866                                                                                 & 0.674                                                                                                      &                          &                         &                                             \\[4ex]

\multirow{10}{*}{0.0500}    & 0.0375                      & 0.999                                                                                 & --                                                                                          & 6,615.0                            & 0.862                                                                                 & 0.670                                                                                                      & \multirow{10}{*}{0.039}  & \multirow{10}{*}{0.010} & \multirow{10}{*}{0.009}                     \\
                            & 0.0425                      & 0.650                                                                                 & --                                                                                          & 14,147.5                           & 0.786                                                                                 & 0.550                                                                                                      &                          &                         &                                             \\
                            & 0.0500                      & --                                                                                    & 0.015                                                                                       & 19,824.5                           & --                                                                                    & --                                                                                                         &                          &                         &                                             \\
                            & 0.0500                      & --                                                                                    & 0.006                                                                                       & 19,935.0                           & --                                                                                    & --                                                                                                         &                          &                         &                                             \\
                            & 0.0500                      & --                                                                                    & 0.006                                                                                       & 19,927.0                           & --                                                                                    & --                                                                                                         &                          &                         &                                             \\
                            & 0.0500                      & --                                                                                    & 0.010                                                                                       & 19,891.5                           & --                                                                                    & --                                                                                                         &                          &                         &                                             \\
                            & 0.0500                      & --                                                                                    & 0.007                                                                                       & 19,907.0                           & --                                                                                    & --                                                                                                         &                          &                         &                                             \\
                            & 0.0575                      & 0.642                                                                                 & --                                                                                          & 14,366.0                           & 0.748                                                                                 & 0.491                                                                                                      &                          &                         &                                             \\
                            & 0.0625                      & 0.992                                                                                 & --                                                                                          & 7,359.5                            & 0.813                                                                                 & 0.601                                                                                                      &                          &                         &                                             \\
                            & 0.0675                      & 1.000                                                                                 & --                                                                                          & 4,372.5                            & 0.858                                                                                 & 0.663                                                                                                      &                          &                         &                                             \\

\end{tabular}%\hspace*{-1cm}
\end{table}
\end{landscape}

\begin{figure}[!ht]
  \caption{Plot of $P(p_{r^{*}} = \max_{r}(p_{r}))$ vs. N; 11-arms, $p_{0}=0.5$}
  \label{pbest_plot}
  \centering
    \includegraphics[width=0.9\textwidth]{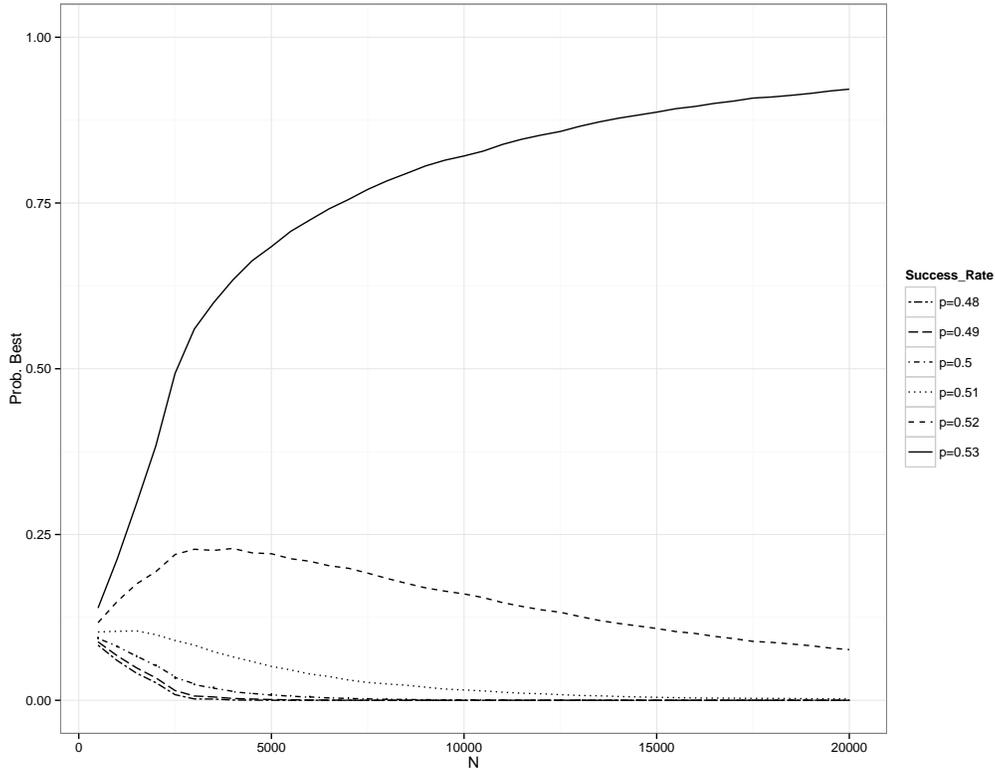}
\end{figure}

For convenience, all five arms with a success rate of 0.5 are condensed into one line in Figure \ref{pbest_plot}. As expected, the arm with a success rate of 0.53, the maximum among all arms in the experiment, breaks away from the rest and converges towards one, while the probabilities of all the other arms being the best slowly fade towards zero. This phenomenon is not unique to this particular simulation, as all simulations presented in Table \ref{table1} and Table \ref{table3} exhibited nearly identical trends. Indeed, this makes intuitive sense. At the beginning of the experiment we don't know anything about any of the arms in the experiment, so \emph{a priori} all arms have an equal probability of being the best. On average, however, as more data is accumulated, we're able to more accurately identify which arm is the best performing arm in the study.
\\

\subsection*{Additional Comparisons}

It is worth noting that the sequential monitoring portion of the method presented in this paper is also quite similar to that of Johari, Pekelis, and Walsh \cite{optimizely} (henceforth referred to as JPW), who appear to have independently derived very similar results to those in \cite{chengshen}. Indeed, they are largely concerned with the same problem presented in this paper - specifically sequential monitoring and multiple comparisons adjustment in A/B tests as it relates to digital marketing. The method in \cite{optimizely}, however, does not carry out model fitting and estimation in a fully Bayesian framework, and employs an FDR correction for multiple comparisons, both of which differ from the method presented here. 
\\

Briefly, JPW \cite{optimizely} (or, in a more accessible version, Pekelis, et al \cite{optimizely_short}) also define their test statistic as the Bayes Factor given by \eqref{eq:bf2}, which they denote $\Lambda_{n}$. They then invert $\Lambda_{n}$ to obtain a p-value, $p^{*}$. For a given Type I error, $\alpha$, $p^{*}$ would be interpreted in the usual way. For a family of p-values, $p^{*}_{i}$, $i = 1, \ldots, m$ obtained from a multiple-testing situation, the authors apply an FDR correction, e.g. \cite{benj_hoch95}, to control the false discovery rate. To compare the method presented here to that of JPW, I applied the method outlined in \cite{optimizely, optimizely_short} to the same 11-arm simulated data that I applied the proposed method to, the results of which were presented in Table \ref{table3}. Table \ref{johari_table} below shows the results when the approach of \cite{optimizely, optimizely_short} is applied to that same data. Specifically, to implement their approach I fit a standard logistic regression model with an indicator for each arm in the experiment as in \eqref{eq:logitm} and calculated the Bayes Factor corresponding to each covariate, $\beta_{r}$, via a BIC approximation. I then took the inverse of each Bayes Factor to obtain the corresponding p-values, and applied the standard Benjamini-Hochberg correction \cite{benj_hoch95} to the p-values at each interim check to control the FDR at 0.05.
\\

It is immediately obvious from Table \ref{johari_table} that the method of JPW, as I have applied it, is extremely conservative when applied to this simulated data. While it's true that the FDR is nearly zero, as is the FWER, in both simulated experiments, this is achieved at the expense of the test's power. This is especially the case when the true difference between arms is small. As the difference between arms increases, the JPW approach begins to perform better, although the average sample size required to detect that difference is still much larger than what is seen from the method I have proposed. Table \ref{comparison_table} provides a side-by-side comparison of the proposed method and the JPW method for a handful of metrics, clearly showing that the proposed method has much higher overall accuracy (a lower overall error rate) and accrues substantially fewer subjects, on average, than does JPW's method. On the other hand, JPW present a method that achieves a lower FDR and a lower false positive rate. 

\newpage

\begin{landscape}
\begin{table}[ht]
\small
\centering
\vspace*{0.18cm}
\caption{11-arm Simulation Results, JPW Method}
\label{johari_table}
\hspace*{0.2cm}\begin{tabular}{ccccrccccc}
\multicolumn{1}{c}{$p_{0}$} & \multicolumn{1}{c}{$p_{r}$}  & \multicolumn{1}{c}{\begin{tabular}{c}Power\\ (1-$\hat{\beta}$)\end{tabular}} & \multicolumn{1}{c}{\begin{tabular}{c}Type I Error\\ ($\hat{\alpha}$)\end{tabular}} & \multicolumn{1}{c}{$\overline{N}$} & \multicolumn{1}{c}{\begin{tabular}{c}Fixed-Sample\\ Avg. Power\end{tabular}} & \multicolumn{1}{c}{\begin{tabular}{c}Fixed-Sample\\ Avg.Power\\ (Bonferroni)\end{tabular}} & \multicolumn{1}{c}{FWER} & \multicolumn{1}{c}{FDR} & \multicolumn{1}{c}{Per-test $\hat{\alpha}$} \\
\hline
\\
\multirow{10}{*}{0.50}      & 0.48                               & 0.311                                                                                 & --                                                                                          & 18,409.0                           & 0.954                                                                                 & 0.829                                                                                                      & \multirow{10}{*}{0.001}  & \multirow{10}{*}{0.001} & \multirow{10}{*}{0.000}                     \\
                            & 0.49                                          & 0.011                                                                                 & --                                                                                          & 19,936.0                           & 0.515                                                                                 & 0.209                                                                                                      &                          &                         &                                             \\
                            & 0.50                                       & --                                                                                    & 0.000                                                                                       & 20,000.0                           & --                                                                                    & --                                                                                                         &                          &                         &                                             \\
                            & 0.50                              & --                                                                                    & 0.000                                                                                       & 20,000.0                           & --                                                                                    & --                                                                                                         &                          &                         &                                             \\
                            & 0.50                                      & --                                                                                    & 0.000                                                                                       & 20,000.0                           & --                                                                                    & --                                                                                                         &                          &                         &                                             \\
                            & 0.50                                      & --                                                                                    & 0.000                                                                                       & 20,000.0                           & --                                                                                    & --                                                                                                         &                          &                         &                                             \\
                            & 0.50                                           & --                                                                                    & 0.001                                                                                       & 19,994.0                           & --                                                                                    & --                                                                                                         &                          &                         &                                             \\
                            & 0.51                                         & 0.013                                                                                 & --                                                                                          & 19,922.0                           & 0.514                                                                                 & 0.209                                                                                                      &                          &                         &                                             \\
                            & 0.52                                        & 0.339                                                                                 & --                                                                                          & 18,168.5                           & 0.953                                                                                 & 0.822                                                                                                      &                          &                         &                                             \\
                            & 0.53                                         & 0.929                                                                                 & --                                                                                          & 11,850.5                            & 0.966                                                                                 & 0.891                                                                                                      &                          &                         &                                             \\[4ex]

\multirow{10}{*}{0.0500}    & 0.0375                      & 0.954                                                                                 & --                                                                                          & 11,258.5                            & 0.966                                                                                 & 0.889                                                                                                      & \multirow{10}{*}{0.001}  & \multirow{10}{*}{0.000} & \multirow{10}{*}{0.000}                     \\
                            & 0.0425                      & 0.204                                                                                 & --                                                                                          & 18,850.0                           & 0.922                                                                                 & 0.736                                                                                                      &                          &                         &                                             \\
                            & 0.0500                      & --                                                                                    & 0.000                                                                                       & 20,000.0                           & --                                                                                    & --                                                                                                         &                          &                         &                                             \\
                            & 0.0500                      & --                                                                                    & 0.001                                                                                       & 19,990.5                           & --                                                                                    & --                                                                                                         &                          &                         &                                             \\
                            & 0.0500                      & --                                                                                    & 0.000                                                                                       & 20,000.0                           & --                                                                                    & --                                                                                                         &                          &                         &                                             \\
                            & 0.0500                      & --                                                                                    & 0.000                                                                                       & 20,000.0                           & --                                                                                    & --                                                                                                         &                          &                         &                                             \\
                            & 0.0500                      & --                                                                                    & 0.000                                                                                       & 20,000.0                           & --                                                                                    & --                                                                                                         &                          &                         &                                             \\
                            & 0.0575                      & 0.163                                                                                 & --                                                                                          & 19,073.5                           & 0.893                                                                                 & 0.666                                                                                                      &                          &                         &                                             \\
                            & 0.0625                      & 0.865                                                                                 & --                                                                                          & 13,319.0                            & 0.961                                                                                 & 0.877                                                                                                      &                          &                         &                                             \\
                            & 0.0675                      & 0.996                                                                                 & --                                                                                          & 8,008.5                            & 0.972                                                                                 & 0.903                                                                                                      &                          &                         &                                             \\

\end{tabular}%\hspace*{-1cm}
\end{table}
\end{landscape}

\newpage

\begin{table}[h]
\small
\centering
\caption{Proposed Method and JPW Method: Simulation Comparison}
\label{comparison_table}
\begin{tabular}{|l|c|c|c|c|}
\hline
                                                                                            & \multicolumn{2}{|c|}{$p_{0} = 0.50$} & \multicolumn{2}{|c|}{$p_{0} = 0.05$}  \\
                                                                                            \hline
                                                                                            & Proposed Method    & JPW Method    & Proposed Method & JPW Method        \\
                                                                                            \hline
FWER                                                                                        & 0.038              & 0.001         & 0.039           & 0.001             \\
\hline
FDR                                                                                         & 0.014              & 0.001         & 0.010           & \textgreater0.001 \\
\hline                                                                                        
\begin{tabular}[c]{@{}l@{}}Overall Error Rate\\ ($(FP + FN)/N_{tests}$)\end{tabular}                  & 20.00\%            & 33.98\%       & 7.61\%          & 18.19\%           \\
\hline
\begin{tabular}[c]{@{}l@{}}Avg. sample accrued for\\ tests where $p_{r} \neq p_{0}$\end{tabular} & 13,684.1            & 17,657.2      & 9,372.1          & 14,101.9           \\
\hline

\end{tabular}
\end{table}

It is possible that some of the difference observed between the two methods is attributable to how I have implemented the JPW method. In particular, in \cite{optimizely, optimizely_short} the authors discuss how an important part of their method involves selecting an ``optimal prior'' for the parameter of interest, something that they base on over 40,000 historical A/B tests run on their company's platform \cite{optimizely}. Still, as that appears to be proprietary information, it is not clear how an independent researcher can adequately evaluate their method without use of anything more than a non-informative prior. Because both methods presented here assume the same prior distribution that is imposed by using the BIC to approximate the Bayes Factor, specifically a unit information prior \cite{raftery99}, I think the comparison is fair. Additionally, anyone wishing to implement an approach like this for the first time or on an ad hoc basis wouldn't have much, if any, historical data to use to tune their prior, so it is important to evaluate each method's performance in the setting with little prior information. Yet, it is also important to consider that the JPW method would likely see improved performance given a well tuned prior for the regression coefficients in the logistic regression. 
\\

Another natural comparison that is not presented here would be this method's performance relative to an experiment with an O'Brien Fleming (OBF) stopping rule \cite{obf}. Cheng and Shen \cite{chengshen} present this comparison in Table 1 of their paper for a simple two-arm test. When the minimum detectable difference is correctly specified, an OBF design will perform just as well as this method with respect to Type I error, power, and average sample size. If, however, the minimum detectable difference is overestimated, the power in an OBF design will take a substantial hit. Alternatively, if the minimum detectable difference is underestimated, the study will be overpowered and result in a longer study that enrolls more people than needed, on average, than the sequential Bayesian design.

\section*{Discussion}

\subsection*{Additional Considerations}

I believe the most natural extension in this framework would be allowing for outcome-adaptive randomization throughout the experiment. That is, as the experiment is ongoing, the study would learn from itself and use the data that has already been accrued to proportionately allocate more people to the best performing arms, while reducing the allocation percentage in arms that are not performing well. Formally, let $$p_{j,r^{*}}(\max)=P(p_{r^{*}} = \max_{r}(p_{r}) \hspace{0.1cm}|\hspace{0.1cm} \mathcal{F}_{j})$$ be the probability that arm $r^{*}$ is the best performing arm at interim check $j$. Then, moving forward, the percentage of study participants to be allocated to any arm $r=r^{*}$ after interim check $j$ can be given by \cite{thall_wathen05, thall_wathen07}:

\begin{equation}\label{eq:adap}
Alloc_{j^{+}}(r^{*}) = \frac{\left\{p_{j,r^{*}}(\max)\right\}^{h}}{\sum_{r}\left\{p_{j,r}(\max)\right\}^{h}},
\end{equation}

where $0 \leq h \leq 1$ is a parameter that adjusts how aggressive the adaptive procedure will be. Setting $h=0$ yields conventional randomization, while $h=1$ gives $Alloc_{j^{+}}(r^{*}) = p_{j,r^{*}}(\max)$. Thall and Wathen \cite{thall_wathen07} note that $h=1/2$ typically works well, while many desirable properties are also obtained setting $h=n/2N$, where $n/N$ is the current sample size divided by the maximum sample size for the experiment, though it could also be thought of as the fraction of the study completed to that point if no maximum sample size has been set. The adaptive randomization could be constructed so that \eqref{eq:adap} does not include the control group, and instead is only calculated for each of the different treatment arms in the study. Such an approach would likely help to hone in on the best performing variant more quickly, but might not be desirable in a setting where the goal is to accurately measure the treatment effect in each arm, because discontinuing allocation to some arms early on in the study should result in more variable estimates in those arms, meaning they'll be pulled closer towards the overall mean across all arms via \eqref{eq:logitpool}. Therefore, depending on the goal of the study, this is potentially a decision the researcher must make before conducting the experiment. More work is required to determine exactly how well an adaptive randomization scheme like \eqref{eq:adap} would perform when used together with this method.

\comment{Separately, in evaluating our method, another natural comparison that we do not present here would be this method's performance relative to an experiment with an O'Brien Fleming (OBF) stopping rule \cite{obf}. Cheng and Shen \cite{chengshen} present this comparison in Table 1 of their paper for a simple two-arm test. When the minimum detectable difference is correctly specified, an OBF design will perform just as well as this method with respect to Type I error, power, and average sample size. If, however, the minimum detectable difference is overestimated, the power in an OBF design will take a substantial hit. Alternatively, if the minimum detectable difference is underestimated, the study will be overpowered and result in longer studies, on average, than the sequential Bayesian design.}

\subsection*{Conclusions}
In this paper I have presented a method for implementing multivariate A/B testing that I believe can be quite useful in online and mobile marketing. The method primarily draws on previous work in clinical trials methodology \cite{chengshen} and applied social sciences research \cite{gelman12} to create a unified approach to both continuous monitoring of A/B tests, and how to adequately account for multiple comparisons in the event of a multivariate A/B test. When compared to other existing methods, the method proposed herein performs quite favorably.
\\

Recall the questions presented on pages one and two of this paper. By approaching the problem from a Bayesian perspective, a researcher can easily rank and compare arms, or variants, to each other by making use of the posterior simulations and calculating $p_{j,r^{*}}(\max)$ for each arm at each interim check, in addition to determining whether each is significantly better than baseline using the sequential monitoring procedure. Additionally, by not specifying the sample size, $N$, in advance, a researcher does not need to worry about setting a minimum detectable difference in the planning stage and thereby potentially enrolling too few or too many people in his study. The method also gives the researcher the freedom to look at the data as often as he pleases, with valid results no matter when he decides to stop the study. Additionally, while the method presented here was done so with binary outcomes in mind, the same rules would apply for a continuous outcome variable.
\\

From the standpoint of an applied researcher, a major advantage of this approach is the relative ease with which it can be carried out. The models to be fitted are familiar, and approximating the Bayes factor using the BIC is straightforward and computationally inexpensive. Using the BIC approximation to the Bayes factor amounts to assuming a unit information prior for all of the parameters in the model, a conservative assumption with respect to rejecting the null hypothesis \cite{raftery99}. Coupled with a conservative rejection criterion of $c_{\alpha/2} = (1-\alpha/2)/(\alpha/2)$, the proposed method can overall be thought of as conservative with respect to Type I error, FWER, and FDR; and that is exactly what is seen in the results presented in Table \ref{table1} and Table \ref{table3}. A valid question about this method is how well it would perform for an experiment with greater than 11 arms - as that is not a scenario considered here. In practice, however, an A/B test comparing more than 10 different variants to a control group is quite rare. Overall, I believe the method would be most useful when a researcher wants to run an experiment with somewhere between two and 20 arms for some pre-specified amount of time, but with no cap on the number of people that can be enrolled in the study over that time period. This falls neatly in line with the idea of running an experimental campaign for some pre-determined number of weeks, with the hope of continuously monitoring the campaign and terminating the experiment once one or two or any arbitrary number of variants show statistically significant lift over baseline. While digital marketing and advertising provide a natural fit for this methodology, it should also work well for an experiment in any field with similar criteria.
\\

%\newpage
\bibliographystyle{unsrt}
\bibliography{bayes_seq_arxiv.bib}
\end{document}